\documentclass[a4paper,]{jpconf}
\usepackage{graphicx}
\usepackage{hyperref,amssymb,subfigure}
\usepackage[square,sort&compress,numbers]{natbib}
\usepackage{iopams}

\def\msun{\,{\rm M_\odot}}

\def\gsim{ \lower .75ex \hbox{$\sim$} \llap{\raise .27ex \hbox{$>$}} }
\def\lsim{ \lower .75ex\hbox{$\sim$} \llap{\raise .27ex \hbox{$<$}} }

\begin{document}
\title{Origin and Implications of high eccentricities in massive black hole binaries at sub-pc scales }

\author{C. Roedig, A. Sesana}
\address{Max-Planck-Institut f{\"u}r Gravitationsphysik, Albert Einstein
Institut, Am M{\"u}hlenberg 1, 14476 Golm, Germany\\
}

\ead{croedig@aei.mpg.de}

\begin{abstract}
We outline the eccentricity evolution of sub-parsec massive black hole binaries (MBHBs)
forming in galaxy mergers. In both stellar and gaseous environments, MBHBs are expected to grow
large orbital eccentricities before they enter the gravitational wave (GW) observational domain. 
We re--visit the predicted eccentricities detectable by space based laser interferometers
(as the proposed ELISA/NGO) for both environments. Close to coalescence, many MBHBs will still 
maintain detectable eccentricities, spanning a broad range from $<10^{-5}$ up to $\lsim0.5$. Stellar
and gas driven dynamics lead to distinct distributions, with the latter favoring 
larger eccentricities. At larger binary separations, when emitted GWs will be
observed by pulsar timing arrays (PTAs), the expected eccentricities are usually quite large,
in the range $0.01-0.7$, which poses an important issue for signal modelling and detection algorithms. 
In this window, large eccentricities also have implications on proposed electromagnetic 
counterparts to the GW signal, which we briefly review.
\end{abstract}

\section{Introduction}
Multimessenger astronomy with massive black hole (MBH) binaries (MBHBs) has been discussed frequently in recent literature
 (e.g. \cite{HolzHughes2005,Armitage:2005,Kocsis2006,Phinney2009,schnittman2011,roedig11,Sesana11,tanaka11}) 
as a unique way of fully understanding the mutual interaction of these systems with their environment. The general 
picture of MBHBs forming after a galaxy merger has been given in \cite{begelman80}, but observational evidence for 
sub--parsec MBHBs is still lacking (see~\cite{colpid2009} for a review). Numerical studies down to parsec
scales \cite{mm01,Escala2005,Dotti07,Dotti2009b} have shown that the two MBHs continue to lose orbital 
energy and angular momentum under the large-scale action of gas/star-dynamical friction, and
end up forming a Keplerian binary. At this point, the gaseous and stellar environment 
furthers the inspiral by either three-body scattering of individual stars and/or the interaction of 
the binary with a circumbinary gaseous disc (e.g. \cite{MerrittReview05,Armitage:2002}). At this late
evolutionary stage, excitation of $e$ was already noticed and studied in both environments 
(see, e.g., \cite{Armitage:2005,Sesana2010}). 

A non zero source eccentricity has important implications for gravitational wave (GW) observations:
on one hand, eccentricity has an impact on source detection and parameter estimation; on the other
hand, measured eccentricities carry valuable information about the dynamical evolution of the detected
systems. In gas rich environments, the surviving detectable eccentricity in the LISA band
($10^{-4}-10^{-1}$Hz) was addressed in \cite{roedig11}, 
stating that $e_{\rm LISA}\sim 10^{-3}-10^{-1}$ depending on the MBHB parameters. However, in view of the recent 
re--design of the LISA mission, estimates have to be re-adapted accordingly. The New Gravitational Observatory 
(ELISA/NGO){\footnote{\url{https://lisa-light.aei.mpg.de/bin/view/}}} will see a shorter portion of the
binary inspiral, with inevitable consequences for parameter estimation, and for residual eccentricity
detectability. Pulsar timing arrays (PTAs) \cite{hobbs10} are expected to detect the first GWs in the low-frequency 
band by the end of this decade. Given their frequency window ($10^{-9}-10^{-7}$Hz),
they will be sensitive to very massive ($>10^8\msun$) systems only , whose chirping is still marginal. 
This implies at the same time that observed MBHBs might still be coupled to their environment. 
We assess here the expected eccentricity distribution for these heavy sources in the PTA band.
We then focus on the possibility of having observable PTA systems that are to some extent still coupled
to either their stellar or gaseous environment, and assess the plausibility of detecting possible EM 
counterparts with current and up--coming missions, such as eROSITA \cite{eRosita10}, MAXI \cite{MAXI09} 
and ATHENA\footnote{\url{http://www.mpe.mpg.de/athena/home.php?lang=en}}. 

The paper is structured as follows: First we describe the mechanisms of eccentricity growth in stellar 
environments in Sec.~\ref{sec:stellar} and gaseous discs  in Sec.~\ref{sec:gas}. We then calculate 
the residual eccentricity distributions in the ELISA/NGO band in Sec.~\ref{sec:ELISA} 
and comment on the expected eccentricity population in the PTA band in Sec.~\ref{sec:PTA}. We review 
the implications this has on multimessenger signals in Sec.~\ref{sec:multi} and 
give our conclusions in Sec.~\ref{sec:conclusions}.

\section{Mechanisms of eccentricity growth:}
In the following we discuss the main physical mechanisms driving the MBHB eccentricity
evolution in the dense astrophysical environment of a post-merger galaxy. We consider 
a MBHB with total mass $M=M_1+M_2$ ($M_2<M_1$ are the individual MBH masses), 
semimajor axis $a$ and eccentricity $e$. Throughout the manuscript, quantities 
without subscripts (such as $M$, $a$, $e$, $E$, $L$) always refer to the MBHB,
whereas subscripts '1' and '2' refer to the individual MBHs.

\subsection{Stellar environment}
\label{sec:stellar}
In dense stellar environments, the MBHB evolution is basically driven by three body
interactions with individual stars \cite{mikkola92,quinlan96,sesana06}. To get a sense of the 
general picture, it is useful to consider the interactions between the stars and the binary 
one by one, as uncorrelated events, even though secular effects may also play a role (see, e.g., \cite{meiron11}). 

At the moment of pairing at sub-pc separations,
the MBHB is expected to be surrounded by a dense stellar distribution of stars. 
If $a$ is the MBHB semimajor axis, three body interactions with stars approaching the 
binary center of mass within $a$ result in superelastic scattering, usually followed by
the ejection of the stars. These stars carry away energy and angular momentum, causing the
binary to shrink. The efficiency of this mechanism depends on several environmental factors,
such as the details of the stellar distribution and the effectiveness of relaxation mechanisms. 
We are here interested in the eccentricity evolution. 
 
Assessing the eccentricity evolution is not straightforward, because it depends on a
combination of energy ($E$) and angular momentum ($L$) exchange during the scattering: 
$\Delta{e}\propto c_i\Delta{E} +c_j\Delta{L}$, where the two $c$ coefficients depend
on the property of the MBHB. Let us consider,
for simplicity, a bound star orbiting the MBHB at a given semimajor axis $a_*$. The 
orbital energy that can be extracted in the scattering is $\propto a_*^{-1}$, whereas the angular
momentum of the orbit is proportional to $\sqrt{a_*}$. It follows that 
$\Delta{e}\propto -c_ia_*^{-1}+c_j\sqrt{a_*}$ \cite{sesana08}. 
If $a_*$ is small enough, then the first term dominates
and the binary circularizes; vice versa, if $a_*$ is large enough, then the second term dominates, and 
the binary eccentricity grows. This behavior has to be expected, since stars with large $a_*$ have large 
angular momentum but just little energy, and the opposite is true for stars with small $a_*$.
It turns out that the transition point is at $a_*\approx a$, i.e., stars within the binary orbit 
on average promote circularization, whereas stars outside the binary orbit make the binary grow
eccentric. Heuristically this can be interpreted as follows. Let us consider a MBHB with 
$M_2\ll M_1$ and a given non-zero eccentricity. Such a binary spends most of
its period near its apocenter, so in the case $a_* > a$ the probability
of a close star-binary encounter (and subsequent star ejection) is
maximal there. In the unequal mass case, the star is ejected following a series of close 
encounters with $M_2$. In such interactions the star gets a small 'kick' and the
velocity of $M_2$ decreases. As this velocity is perpendicular to the
MBH position vector close to the apocenter, the secondary is forced on a more 
radial orbit. On the contrary, a star with $a_* < a$ "feels" the secondary
hole when this approaches the primary at a distance $<a$.
At that point, the interaction with the star is unlikely to occur close
to the pericenter of the MBHB orbit (because the time spent by the binary at the pericenter
is very small), and typically also extracts a large radial
component from the velocity of the secondary black hole, hence
causing circularization. Considering the collective effect of the stellar
distribution, the evolution of the orbital eccentricity is therefore 
governed by the relative weight of stars inside and outside $a$. The steeper
is the stellar cusp, the less effective is the eccentricity growth.  
As a general trend, quasi circular-equal mass MBHBs experience just a mild
eccentricity growth, while systems which are already eccentric
at the moment of pairing, or with mass ratio significantly lower than
1, can evolve up to $e>0.9$ \cite{Sesana2010}. Such a trend has been observed in
several numerical studies \cite{mm01,mms07,mat07,as09}.
    
The above argument is strictly valid for isotropic stellar distributions.
The evolution of the binary eccentricity can be extremely different
for non isotropic systems. For example, \cite{sgd11}
demonstrated that in a rotating stellar system, the eccentricity
evolution of unequal MBHBs is dramatically affected by the level of
co/counter rotation of the stellar distribution with respect to the
binary, with corotating distributions promoting circularization rather
than eccentricity growth. This is because, as a simple consequence of 
angular momentum conservation during the ejection process, 
stars counterrotating with the binary tend to extract a lot of angular 
momentum from the MBHB, causing the eccentricity growth, whereas 
corotating stars do not. Nonetheless, most of the simulations
involving rotating bulges \cite{ber06,Berentzen2009}, or merging systems
\cite{khan11,preto11} find quite eccentric binaries at the
moment of pairing (ranging from $0.4$ to $0.8$), and the subsequent
evolution leads to a general eccentricity growth, in good agreement of
what predicted for an isotropic stellar distribution. This may be
because the binary evolution is mostly driven by loss cone refilling
of unbound stars on almost radial orbits, with negligible initial
angular momentum. On the other hand, \cite{mg11} find milder eccentricity
evolutions, along the lines of what is expected for a rotating stellar
remnant. The eccentricity evolution of MBHBs in stellar environments
is still a largely open issue, but a significant eccentricity seem to be a general 
outcome of the models.

\subsection{Gaseous environments}
\label{sec:gas}
The study of the secular evolution of a MBHB immersed in a gaseous disc is essentially similar to
type-II planet migration \cite{GoldreichSari03}. However, when self-gravity of the disc 
and the accretion of matter onto the MBHs is included,
the clear predictions of resonances and decay times coming from perturbation theory tend to be 
smeared out. In this paragraph, we will sketch the two effects that lead to the excitation and 
the saturation of the eccentricity evolution:
\begin{itemize}
 \item excitation of eccentricity via asymmetric torques from the circumbinary disc 
 \item saturation of eccentricity growth due to accumulated gas around the black holes
\end{itemize}
When dynamical friction becomes inefficient, the subsequent MBHB evolution relies on
transfer and redistribution of energy and angular momentum from the binary to the 
so--called circumbinary disc.
This usually occurs at sub-parsec binary orbital separations, and goes hand in hand with the 
appearance of a so--called \textit{gap}. A \textit{gap} is a hollow, low--density region which
is found to have a radial extent $\delta$ of about twice the semi-major axis $a$ of the binary 
(e.g. \cite{MacFadyen2008,Jorge09,roedig11}) which is attributed to the excitation of outer
Lindblad resonances (OLR) with the most prominent one at $\delta_{m=1}=2^{2/3}a \sim 1.59\,a$
\footnote{The excitation of the $m=1$ OLR is suppressed in equal mass binaries, nevertheless 
similar gap-sizes are also found for exactly equal mass MBHBs \cite{MacFadyen2008}}.
However, gas will still continue to leak from the inner edge of the disc
to the black holes and form \textit{mini-discs} inside the respective Roche-lobes 
\cite{arty96,hayasaki07,roedig11,Sesana11}. As first shown by \cite{Armitage:2005} and confirmed by 
\cite{Jorge09}, a binary inside a gap will not remain circular and neither will
the inner edge of the disc (see also \cite{Lubow1991,Papaloizou2001,MacFadyen2008}).\\
As showed in \cite{roedig11} for a MBHB of total mass $M$ and mass-ratio $q= M_2/M_1=1/3$,
 the eccentricity of the binary will continue to grow up to about $e\sim 0.6$ while
the eccentricity of the disc stays small, $e_{\rm disc} < 0.15$, even for $e >0.7$. 
The reason for this growth of $e$ can be understood by comparing the angular
velocity of the black holes to that of the fluid elements in the circumbinary disc.
The movement of the MBHB inside the disc will induce overdensities close to the inner rim of 
the circumbinary disc, moving at lower angular speeds than the MBHs.
If the MBHs are at apocenter, where they are slowest and thus remain the longest,
they will experience a deceleration (negative torque) thus decreasing their 
associated angular momenta $L_{1,2}$ and increasing $e$. 
Since the disc is at the same time expanding radially to accommodate the larger
maximum MBHB separation, the theoretical limit to this growth depends on the evolution of 
$\delta(e,t)$. Equating the average angular velocity of the inner rim of the disc to that of the binary 
at apocenter: $\langle \omega_{\rm fluid} \rangle= \omega_{\rm apo}$ yields the implicit relation for a limiting $e$:
\begin{equation}
\frac{(1+e)^3 }{(1-e)}=\delta_{a}(e,t)^3,\,\,\,\,\,\,\,\,\,\delta_{a}=\delta/a
\label{eq:limit1}
\end{equation}
which shifts the question to determining the function $\delta(e,t)$. 
Assuming $\delta(e,t)=const=2$, \cite{roedig11} found a limiting eccentricity of $\approx 0.6$. However
we expect $\delta(e,t)$ to generally be a growing 
function of $e$ related to the strongest OLRs. A detailed discussion of the OLRs and their evolution will soon
be found in \cite{roediginprep}. In general, if $\delta$ grows with $e$, $L_{\rm fluid} < L_{2}$ is always 
true, and $e$ continues to grow unless some other effect comes into play.
When the binary is very eccentric, the Roche-lobes and the mini--discs attached to each individual MBH
overlap at pericenter. Whenever this starts happening the secondary MBH will have to pass through 
the mini--disc of the primary and it will be slowed down at pericenter, decreasing $e$. This effect 
is thus strongly dependent on the thermodynamics of the gaseous disc and the mass contained within 
the mini--disc.

\begin{figure}
\centering
\includegraphics[width=0.9\linewidth]{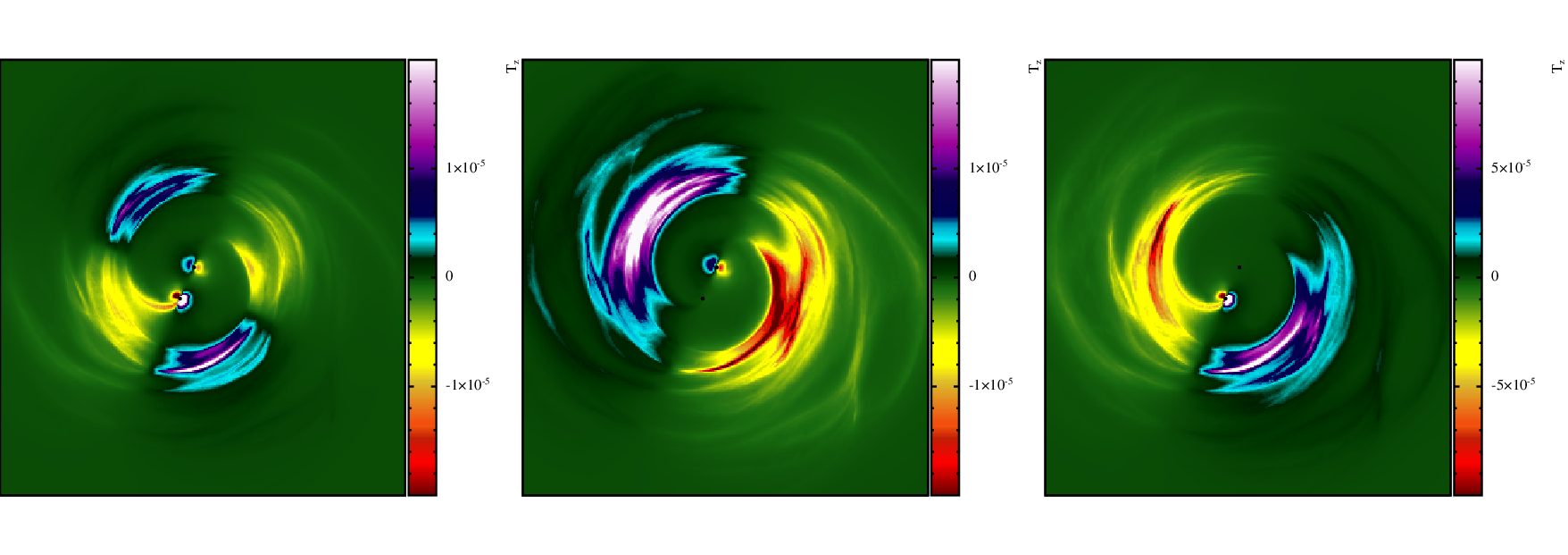}\\
\includegraphics[width=0.9\linewidth]{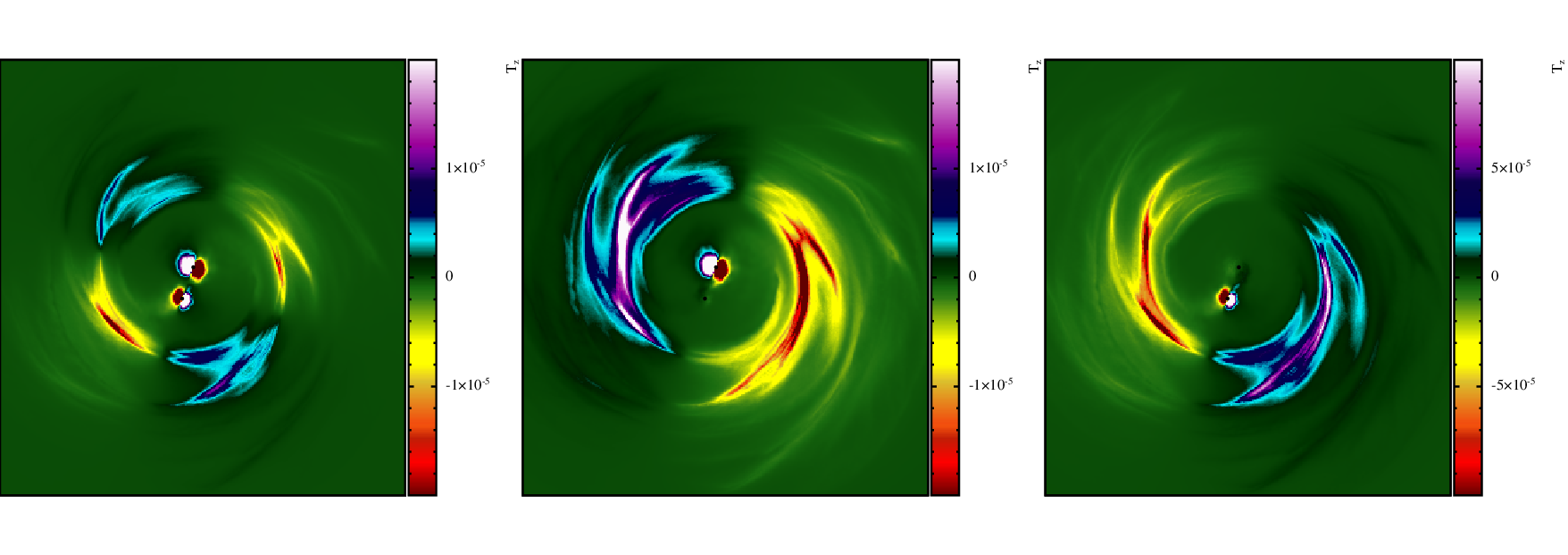}
\caption{Torque density in the $z$-direction  for two initially circular MBHB runs reaching
$e\approx0.1$ after $90$ orbits. Top (bottom) panels: \textit{dry} (\textit{wet}) 
thermodynamic prescription for the cavity. In each line, the three panels are, from left to right: 
the total torque density on the binary  $T_z= M_1 \, T_{z_1}+ M_2 \, T_{z_2}$,
 torque onto the primary MBH $T_{z_1}$, torque onto the secondary MBH $ T_{z_2}$. 
Figure made using {\sc SPLASH}\cite{Price2007}}
\label{fig:torque}
\end{figure}

In the remainder of this section, we will elucidate the growth of $e$ with numerical examples of Newtonian 
self-gravitating discs around a MBHB with $q=1/3$, using a setup described in \cite{roedig11}, the details 
of which will soon be found in \cite{roediginprep}.
The two examples use the same initial data and code, differing only in the thermodynamics
treatment inside the gap. The \textit{dry} run allows for an adiabatic evolution of the internal 
energy plus a $\beta$ cooling inside the gap which effectively suppresses gas-inflows, whereas 
the \textit{wet} run furthers the inflows by treating the gas isothermally when crossing the
radius of $r_{\rm iso}=1.73\,a$, thus allowing us to differentiate the effects of the gas in the gap. 
When plotting the torque density $T$ exerted by the disc onto the MBHs in figure~\ref{fig:torque}, 
we show the individual torques in the $z$-direction onto each MBH 
in the middle and right panels and the sum  $T_z= M_1 \, T_{z_1}+ M_2 \, T_{z_2}$ in the left panel.
Since the binary is rotating counterclockwise, the negative portion of the torques always stems 
from the upstream region (yellow/red).  Comparing the top and the bottom set of panels,
we observe that altering the prescription for the thermodynamics inside the gap alters the torque density 
and the dynamical evolution of the entire disc. It is clear that the torque
from the mini--discs are larger for the \textit{wet} run (bottom panels), since the mass inside the mini--discs 
is higher and the gas is not allowed to cool (i.e. its internal energy remains constant) after
crossing the threshold radius $r_{\rm iso}$. The torque density onto $M_1$ (middle panel) is more or less 
symmetric in the azimuthal direction, whereas $M_2$ clearly experiences larger torques from 
the part of the disc that is close to it. What is moreover notable, is the fact that in the left panel, where
the total torque is shown to have a quadrupolar structure,
the contributions from the mini--discs do not have exactly the same orientation as the contributions
from the disc. However, the growth of $e$ is very similar for \textit{wet} and \textit{dry} runs, which leads us to 
conclude that the \textit{excitation} of $e$ stems mostly from the circumbinary disc and is robust under 
different numerical prescriptions. The precise value of $e_{\rm limit}$, however, depends on the density 
and angular momentum of the mini--discs, but is expected to be $e_{\rm limit}\in [0.6,0.8]$. 
Further studies of the precise dependence on the thermodynamics are ongoing \cite{roediginprep}.  

\section{Implications for ELISA/NGO and PTA sources}
\label{sec:implications}
Low-frequency GWs are expected to be rich in astrophysical information if the source parameter 
analysis is sufficiently accurate to differentiate between various proposed astrophysical scenarios. 
Great hopes lie in a coincident detection of EM counterparts to complement
the GW information with an independent measurement of the source redshift and a connection with 
the host galaxy, opening new scenarios for precision cosmology (see, e.g, \cite{HolzHughes2005}).
 The proposed ELISA/NGO instrument is designed for $10^3-10^7\msun$ MBHBs 
chirping and merging in the $10^{-4}-10^{-1}$ Hz band, 
whereas PTAs will be sensitive to GW frequencies in the range $10^{-9}-10^{-6}$Hz with 
successful detections  depending on the sensitivity of the radio instruments, the number of 
available stable millisecond pulsars and the observation time.\\
To guide the development of effective detection pipelines, it is reasonable to review the assumptions 
of GW sources emitting purely sinusoidal, monochromatic signals. To estimate the impact of the eccentricity
growth via the mechanisms described above, we need to understand at which binary separation they are effective.
We therefore compare the star/gas driven shrinking timescales to the GW energy loss timescale.
The GW timescale is given by:
\begin{equation}
t_{\rm GW}=7.84\times10^7 \,{\rm yr}\, M_8 q_{\rm s}^{-1} a_3^{4} F(e)^{-1},\,\,\,\,\,\,F(e)=(1-e^2)^{-7/2}\left(1+\frac{73}{24}e^2 +\frac{37}{96}e^4 \right)
\label{tgw}
\end{equation}
where $q_{\rm s}=4q/(1+q^2)$ is the symmetric binary mass
ratio, $M_8=M/10^8\msun$ and $a_3$ is the binary 
separation in units of $10^3R_S$ ($R_S=2GM/c^2$).

In gas rich environments, assuming a $\beta$--disc model \cite{ss73} for the circumbinary disc, the 
migration timescale is  \cite{Sesana11}
\begin{equation}
t_m=2.09\times10^6 \,{\rm yr}\, \alpha_{0.3}^{-1/2}\left(\frac{\dot{m}_{0.3}}{\epsilon_{0.1}}\right)^{-5/8}M_8^{3/4} q_{\rm s}^{3/8}\delta_a(e)^{7/8} a_3^{7/8}.
\label{tm}
\end{equation}
Here we defined the viscosity parameter $\alpha_{0.3}=\alpha/0.3$, the accretion efficiency 
$\epsilon_{0.1}=\epsilon/0.1$ and the mass accretion rate $\dot{m}_{0.3}=\dot{m}/0.3$, 
where $\dot{m}=\dot{M}/\dot{M}_{\rm Edd}$ is the accretion rate normalized to the Eddington rate.

In a star dominated system, assuming a full loss cone \cite{Sesana2010} the MBHB hardening timescale 
is instead given by
\begin{equation} 
t_{h}= 2.89\times10^6 \,{\rm yr}\, \sigma_{100} M_8^{-1}\rho_5^{-1} a_3^{-1} H_{15}^{-1},
\end{equation}
where we introduced the stellar velocity dispersion $\sigma_{100}=\sigma/100$ Km s$^{-1}$, the stellar 
density at the binary influence radius (see \cite{Sesana2010}, for details) $\rho_5=\rho/10^5\msun$ 
pc$^{-3}$ and the dimensionless hardening rate $H_{15}=H/15$. Note that the combined $M_8^{-1}$ and  
$a_3^{-1}$ scaling implies that $t_h$ is independent on the MBHB mass.

By equating $t_m$ and $t_h$ to $t_{\rm gw}$, one gets the characteristic $a$ at which GW emission takes over 
in the binary evolution. It turns out that binaries are still coupled to their environment down to a 
separation of few hundred Schwarzschild radii, where the most massive systems fall in the PTA frequency band. 
This has two direct consequences: (i)
PTA sources are generally eccentric, with an eccentricity distribution directly determined by the shrinking mechanism, 
and (ii) significant residual eccentricity might be present close to coalescence, in the ELISA/NGO band. 

In the following we apply selfconsistent models for the evolution of the MBHBs in gaseous and stellar environments
to the expected population detectable by LISA/NGO and PTAs. In gas rich systems, we assume MBHBs to maintain
an eccentricity of $e_{\rm limit}=0.6$ as long as they are coupled to their environment and then to circularize
under the action of GW emission after decoupling. In star dominated systems, we use the self--contained model
for the evolution of the MBHB under the effect of star ejection and GW radiation as outlined in \cite{Sesana2010}. 

\subsection{Eccentricity population of ELISA/NGO sources}
\label{sec:ELISA}
\begin{figure}
\includegraphics[width=84.0mm]{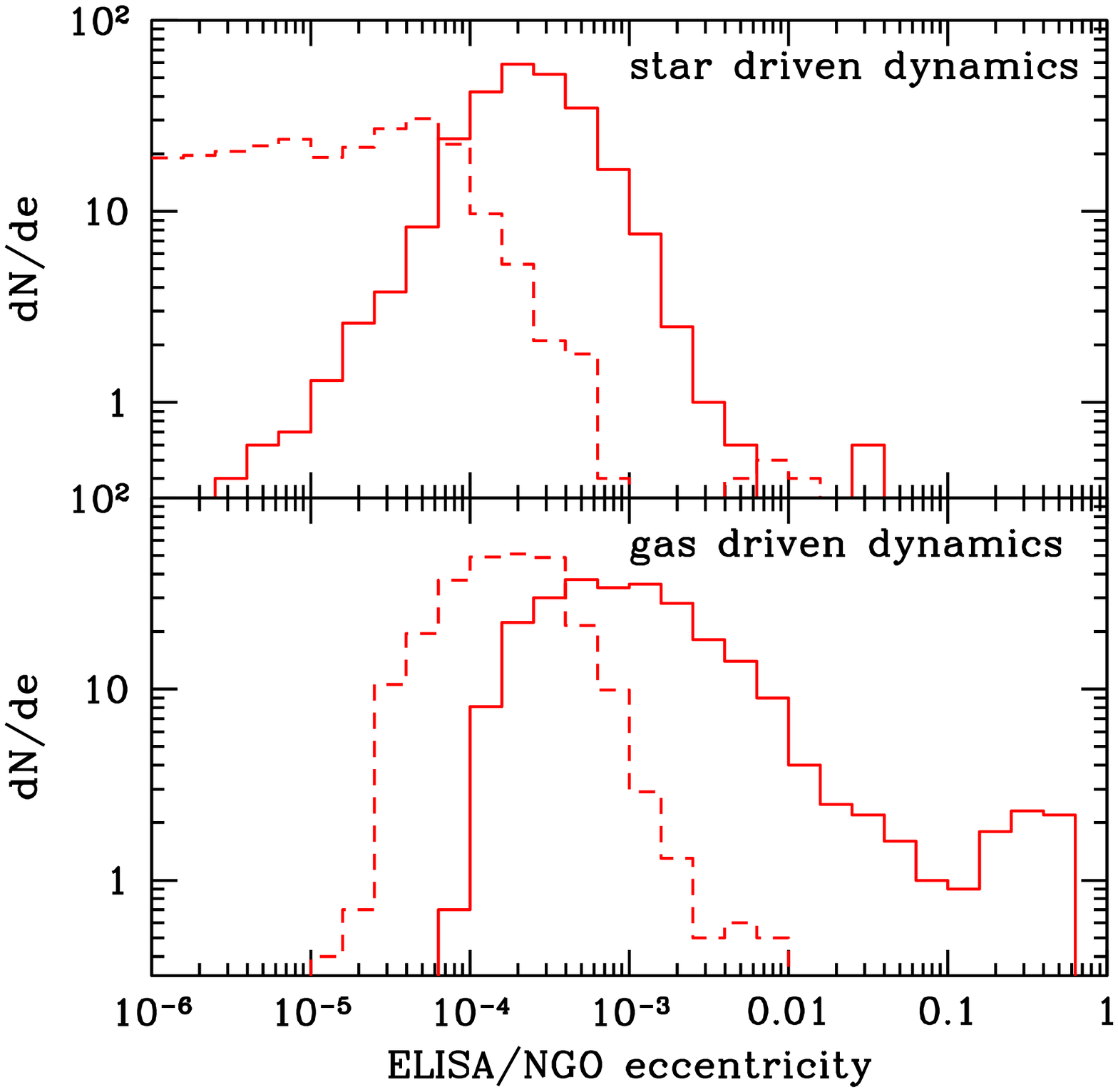}
\includegraphics[width=84.0mm]{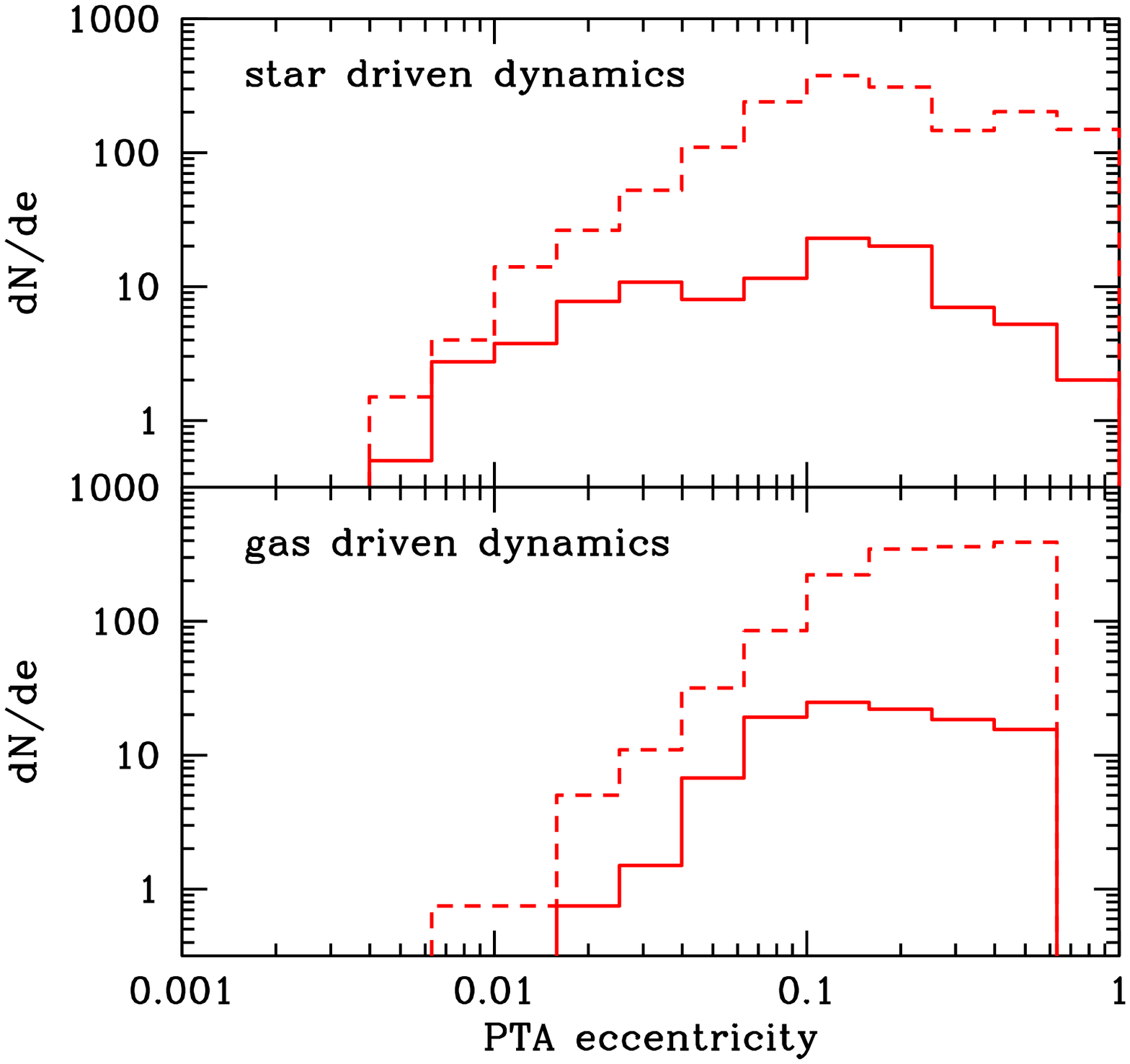}
\caption{Eccentricity population of MBHBs detectable by ELISA/NGO and PTAs, expected in stellar 
and gaseous environments. Left panel: The \textit{solid } histograms represent the efficient 
models whereas the \textit{dashed} histograms are for the inefficient models. Right panel: 
\textit{solid} histograms include all sources producing timing residuals above 3 ns, 
\textit{dashed} histograms include all sources producing residual above 10 ns.}
\label{fig:eccdist}
\end{figure}

In view of the new sensitivity curve of the ELISA/NGO instrument, we re--visit here the
question of the residual detectable eccentricity of a population of MBHBs evolving either 
in stellar or gaseous environments. We take the population of merging MBHBs predicted
by a standard MBH cosmic evolution model where MBHs grow by merger and accretion starting
from a population of light seed MBHs, thought to be remnants of the PopIII stars \cite{Madau2001}.
To be specific, we took the model "LE" from \cite{arun09}. We stress here that in this model, MBHBs were 
assumed to instantly coalesce after a dynamical friction timescale from the pairing of the 
two galaxies hosting the two MBHs: none of the dynamical evolution models is implemented
selfconsistently in deriving the MBHB population. We therefore have a single MBHB population
to which we apply different dynamical evolution models to determine the residual eccentricity.
For each mechanism (gas/star) we consider two scenarios (efficient/inefficient), to give an 
idea of the expected eccentricity range. The models are the following
\begin{enumerate}
\item gas-efficient: $\alpha=0.3$, $\dot{m}=1$. The migration timescale is maximized for this high
values of the disc parameters, and the decoupling occurs in the very late stage of the MBHB evolution;
\item gas-inefficient: $\alpha=0.1$, $\dot{m}=0.1$. Decoupling occurs at much larger (factor 3-to-5)
separations, and MBHBs have much more room to circularize;
\item stars-efficient: $p(e_i)\propto e$. Initial eccentricities are taken according to a thermal 
probability density function \cite{preto11};
\item stars-inefficient: $e_i=0$. Binaries are initially circular, a condition that minimizes the 
effectiveness of eccentricity growth.
\end{enumerate}
Results are shown in the left panel of figure~\ref{fig:eccdist}, where we plot the residual eccentricity of all
sources detected with SNR$>8$, when they enter in the ELISA/NGO band{\footnote{The band entrance
is defined as the frequency where the strain of the source $h_s$ squared equals the one-sided noise
spectral density of the detector $S(f)$.}}. In general, MBHBs evolving in gaseous environments are
expected to retain a larger eccentricity, with a distribution tail possibly extending to $e\sim0.5$.
Residual eccentricities are mostly in the $10^{-4}-10^{-2}$ range, and are generally lower for
star driven systems. Note that the distribution predicted by the star-efficient model just
overlaps to the one predicted by the gas-inefficient one. Therefore, apart from those two extreme cases, 
the distributions predicted by the two families of models are generally distinct, especially at the high $e$ tail. 
A sufficient source detection statistics should allow discrimination between gas and star driven
dynamics, providing valuable information of the MBHB environment.

\subsection{Eccentricity population of PTA sources}
\label{sec:PTA}
To investigate PTA sources, we instead use  MBHB population models from \cite{ks11}. Here we just want to highlight
the fact that PTA sources might be very eccentric, and adequate signal processing algorithms need to be developed.
We therefore use a single representative model for each mechanism. For the gas driven evolution we 
use $\alpha=0.3$, $\dot{m}=0.3$; whereas for the star driven model we use $e_i=0.1$ for all the binaries.
In the right panel of figure~\ref{fig:eccdist}, we plot the eccentricity distribution of all sources contributing
to the signal at timing residual level above $10$ns or $3$ns, which is expected to be in the reach of future 
PTA campaigns. The distributions are rather similar with more spread and higher extremal eccentricities
for the stellar case. It is not clear, whether the two distributions are discernible; the figures, 
however, clearly suggest that a large portion of PTA signals would  have fairly high eccentricities
($e>0.1$). 

\paragraph{Multimessenger signals.}
\label{sec:multi}
Concentrating on gaseous disc environments, \cite{Sesana11} and \cite{tanaka11} have, in a complementary 
way, outlined possible counterparts to resolvable PTA sources, which we will quickly review here. First, 
we will look at coupled systems, as done in \cite{Sesana11}, then
conclude with a remark on how eccentricity impacts on decoupled systems as studied in \cite{tanaka11}.\\
If $t_{\nu}$ is the viscous time of the disc, then as long as $t_{GW}>t_{\nu}$, the disc can dissipate 
inwards fast enough to follow the MBHB and during this phase the MBHB is expected to be highly eccentric.
In \cite{roedig11}, it has been shown that the strength of a periodicity in the light curve related to the
MBHB orbital frequency depends on the eccentricity of the orbit and the angular momentum 
distribution of the inspiralling gas. First, it is important to note that for circular or mildly
eccentric orbits the periodicity in the gas inflow rate is much weaker than for high
($e>$0.4) eccentric systems. 
The number and quality of observations needed to identify the orbital periodicity 
is challenging for upcoming all-sky surveys, even assuming highly modulated inflow rates 
for $e\sim 0.6$ systems. This is simply because of the high number of 
pointings per orbit needed (which will be the main limitation of eROSITA) and the required flux limit 
of the instrument (which limits ongoing surveys like MAXI). 
Detecting a PTA counterpart in  a gaseous environment will thus be likely biased towards eccentric 
systems that accrete the instreaming gas on time scales comparable to the binary period. 
If, however, in the above scenario the time scale for the accretion of the mini--disc is long compared 
to the binary period, the instreaming gas periodicity might be diluted in the process
of angular momentum redistribution within the disk, and might not be reflected 
in the accretion luminosity onto the two MBHs.
In this case, stable, broad double $K \alpha$ lines arising from the inner portion of the two mini--discs
might be detectable by ATHENA \cite{Sesana11}, and could provide meaningful measurements of the
two MBH spins. If, on the contrary, the gas is instantaneously accreted, there will be no detectable 
lines and the general contribution from of gap region to the emitted luminosity will likely be negligible; 
only if the spread in inflowing angular momentum
is very large, colliding caustics can form small inviscid accretion discs~\cite{beloborodov2001,zalamea2009} 
making the MBHs variable, luminous X--Ray sources. 

Looking at circular binaries, \cite{tanaka11} have shown that gas can continue
 to leak towards the MBHB even after decoupling, finding surface densities as large as
$\Sigma\sim 10^4 \, {\rm g/cm^2}$ at $\sim 40 \,GM/c^2$ for a total MBHB mass of $10^9 \msun$. While for eccentric
systems the width of the gap is generally larger than for circular ones,
their Green function approach is valid nonetheless, as long as the leaking
gas is ripped out of the edge of the circumbinary disc on dominantly radial
orbits, i.e. having a much lower angular momentum than the inner edge of the
circumbinary disc.  Their results (see, e.g. their figure~4) would not change  
much qualitatively, if they instead used a larger cavity to begin with and 
then adapted the inner boundary condition to a function $\delta_{a}(e(t))$ 
rather than using $2\, a(t)$ . 
However, due to gas from the decoupled disc being required 
to diffuse over larger distances while the gas is also colder at the same time thus less diffusive, it is expected that the total
amount of fossil gas is smaller the higher the eccentricity.

\section{Conclusions}
\label{sec:conclusions}
We have shown that the excitation of MBHB eccentricity in both star and gas dominated 
environments is a robust prediction of state of the art dynamical evolution models. In 
the gas dominated case, limiting eccentricity values $e_{\rm limit} \ge 0.6$ are expected 
following excitation by a cold dense circumbinary disk, whereas stellar dynamics can easily produce 
eccentricities higher than 0.9. As a consequence, PTA sources, many of which are at the evolutionary
stage in which the environment still plays a role in their dynamical evolution, are expected to be
significantly eccentric. Many binaries producing residuals at a $10$ns level (in the reach of future PTA 
campaigns) have $e>0.1$. The impact of such high eccentricities on the PTA signal should be 
appropriately taken into account when developing dedicated data analysis algorithms.
For ELISA/NGO sources, which have long since decoupled from their environment, we have shown the 
distribution of the residual eccentricity expected for both environments.
As the latter are disjoint at the high-eccentricity tail, enough  GW observation statistics 
could distinguish among the different scenarios, providing valuable information on
the environment of coalescing MBHBs. As far as multimessenger signals are concerned, 
we did not treat electromagnetic counterparts to ELISA/NGO events and focused on PTA observations.
In this latter case, the electromagnetic signal is generally bright, due to the high mass and 
relatively small luminosity distance of the sources, making certain distinct features
detectable by upcoming observatories. We have shown that counterpart identification is likely to
be bias towards eccentric MBHB in periodicity searches, whereas, depending on the angular
momentum of the instreaming material feeding the mini--discs, peculiar double $K\alpha$ lines
might be identifiable by future hard X--ray deep spectroscopy.

\ack
The authors thank Takamitsu Tanaka for helpful conversations and Massimo Dotti for carefully reading the manuscript and giving detailed comments.

\bibliographystyle{iopart-num}
\bibliography{biblio/aeireferences,biblio/ref2}
\end{document}